\documentclass[preprint,
superscriptaddress,
amsmath,amssymb,aps,showkeys,showpacs,
twoside,final,secnumarabic,
nofootinbib]{revtex4-2}

\usepackage[paperwidth=205mm,paperheight=290mm,top=17mm,bottom=25mm,
inner=17mm,outer=17mm,
twoside]{geometry}

\usepackage{cmap} 
\usepackage[T1,T2A]{fontenc}
\usepackage[utf8]{inputenc}
\usepackage[russian,english]{babel}
\usepackage{color}
\usepackage{graphicx}
\usepackage{dcolumn}
\usepackage{bm} 
\usepackage[unicode=true,colorlinks=true,linkcolor=magenta, urlcolor=blue, citecolor = blue,breaklinks]{hyperref}
\usepackage{multirow}
\usepackage{url}
\usepackage{breakurl}
\DeclareGraphicsExtensions{.eps,.png,.pdf}

\newcount\issue
\newcount\Vol
\newcount\numb
\usepackage{fancyhdr} 
\def\Vol{\textbf{80}}
\def\numb{x}
\begin{document}

\title{Deep learning methods for gamma event selection in\\TAIGA-IACT image analysis in stereo-mode} 

\def\addressa{Irkutsk State University, Institute of Applied Physics, Irkutsk, 664003, blv. Gagarina, 20, Russia}
\def\addressb{Lomonosov Moscow State University, Skobeltsyn Institute of Nuclear Physics, Moscow, 119991 Leninskie gory,1, bld.2, Russia}
\author{\firstname{E.O.}~\surname{Gres}}
\email[E-mail: ]{greseo@mail.ru }
\affiliation{\addressa}
\affiliation{\addressb}
\author{\firstname{A.P.}~\surname{Kryukov}}
\affiliation{\addressb}
\author{\firstname{P.A.}~\surname{Volchugov}}
\affiliation{\addressb}
\author{\firstname{A.P.}~\surname{Demichev}}
\affiliation{\addressb}
\author{\firstname{J.J.}~\surname{Dubenskaya}}
\affiliation{\addressb}
\author{\firstname{D.P.}~\surname{Zhurov}}
\affiliation{\addressa}
\affiliation{\addressb}
\author{\firstname{S.P.}~\surname{Polyakov}}
\affiliation{\addressb}
\author{\firstname{E.B.}~\surname{Postnikov}}
\affiliation{\addressb}
\author{\firstname{A.J.}~\surname{Razumov}}
\affiliation{\addressb}

\begin{abstract}
In astrophysical complex TAIGA there are four Imaging Atmospheric Cherenkov telescopes (IACTs), located at a distance of 300-500 m from each other, 
due to which extensive air showers (EAS) generated by cosmic and gamma radiation are recorded by a single telescope or by multiple telescopes simultaneously (stereo-mode). 
The analysis of EAS images from two or more Cherenkov telescopes together is of particular interest, since for this approach the reconstruction of physical parameters of the primary 
gamma ray is much more accurate than in mono-observations. However, the gamma-ray event reconstruction strongly depends on the quality of the hadron 
background suppression of the classifier, therefore, the gamma event selection stage is one of the important ones in the further 
analysis of IACT data. We propose and study several neural network methods for gamma event selection in joint TAIGA-IACT image analysis. Neural network methods we suggested 
involves a single-stage analysis of IACT images from two telescopes, while the second method considers the 
formation of autoencoder-derived essential features that contain the EAS physical parameters, including the type of primary particle.
\end{abstract}
\par
\keywords{IACT, stereo-mode, gamma/hadron selection, convolutional neural network methods\\[5pt]}

\maketitle
\thispagestyle{fancy}

\section{Introduction}\label{intro}

The study of sources of gamma radiation of high and very high energy is of great interest to astrophysics and gamma-ray astronomy. Since gamma quanta have no charge, the gamma radiation source determines clearly on the celestial sphere. In combination with other methods of multimessenger astronomy, it allows astrophysicists to obtain the full picture of any astrophysical object. In very high energy gamma astronomy, detection of gamma radiation is only possible from the Earth surface, implementing extensive air shower registration\cite{1, 2}.

Extensive air shower (EAS) is a developing cascade of secondary particles that occurs when the primary particle interacts with an atmosphere atom. The primary particle can be a gamma ray or the atomic nucleus from the cosmos. Depending on the type of the primary particle, EAS development differs in form and composition. Based on this knowledge, various methods and techniques are being developed for ground-based installations and experiments in order to separate the gamma ray flux from hadron flux in cosmic radiation as efficiently as possible. The aim of our work is to develop and adapt neural network methods to the gamma/hadron separation problem using data obtained by atmospheric Cherenkov telescopes of the TAIGA complex.

\section{TAIGA-IACT}

On the territory of the gamma-observatory TAIGA there are several types of detectors that register the Cherenkov radiation produced by atmospheric showers\cite{3, 4}. Four Imaging Atmospheric Cherenkov Telescopes of TAIGA (TAIGA-IACT) are one of the detectors of this complex in the task of separating gamma radiation from cosmic background radiation. Each telescope has a composite mirror of the Davis-Cotton optical
system with a focal length of 4.75 m. In the focus of the mirrors, camera of 600 photomultipliers (XP1911) with a photocathode diameter of 2 cm is installed. The field of view of camera is $9.6^\circ$, the field of view of one pixel is $0.36^\circ$.

A special feature of various IACT experiments (MAGIC, HESS, CTA, etc.), including TAIGA-IACT, is that the telescopes are placed close to each other. As a result, the Cherenkov light cone can trigger one, two or more telescopes. The analysis of the image from one telecope is called monoscopic observation mode (mono-mode). Meanwhile, the method of analyzing multiple images from different points is called stereoscopic observation mode (stereo-mode). The minimum energy of primary particle required for detecting EAS with several telescopes is slightly higher compared to monoscopic observations with Cherenkov telescopes. Thus, the TAIGA-IACT registers events with an energy starting from 3 TeV in mono mode, while telescopes in  stereo mode register events with an energy of 10 TeV and higher.

Currently, the traditional TAIGA-IACT  image processing approach in stereo mode is based on the Hillas parameter method\cite{5}. In particular, each image is parameterized by an ellipse based on the principal component analysis method, while the position of the source is approximated using Monte Carlo simulation. Applying and optimizing cuts on the shape, location and rotation of the ellipse on the image, taking into account multiple telescopes, gamma-like events are selected.

Since the traditional approach uses only a generalization of image calculated as the central statistical moments of images (no more than the fourth moment, so a lot of information may be lost during the analysis), in this article the possibility of using neural network methods in image processing is explored.

\section{Neural network techniques proposed for TAIGA-IACT stereo-mode}

\begin{figure*}
\includegraphics{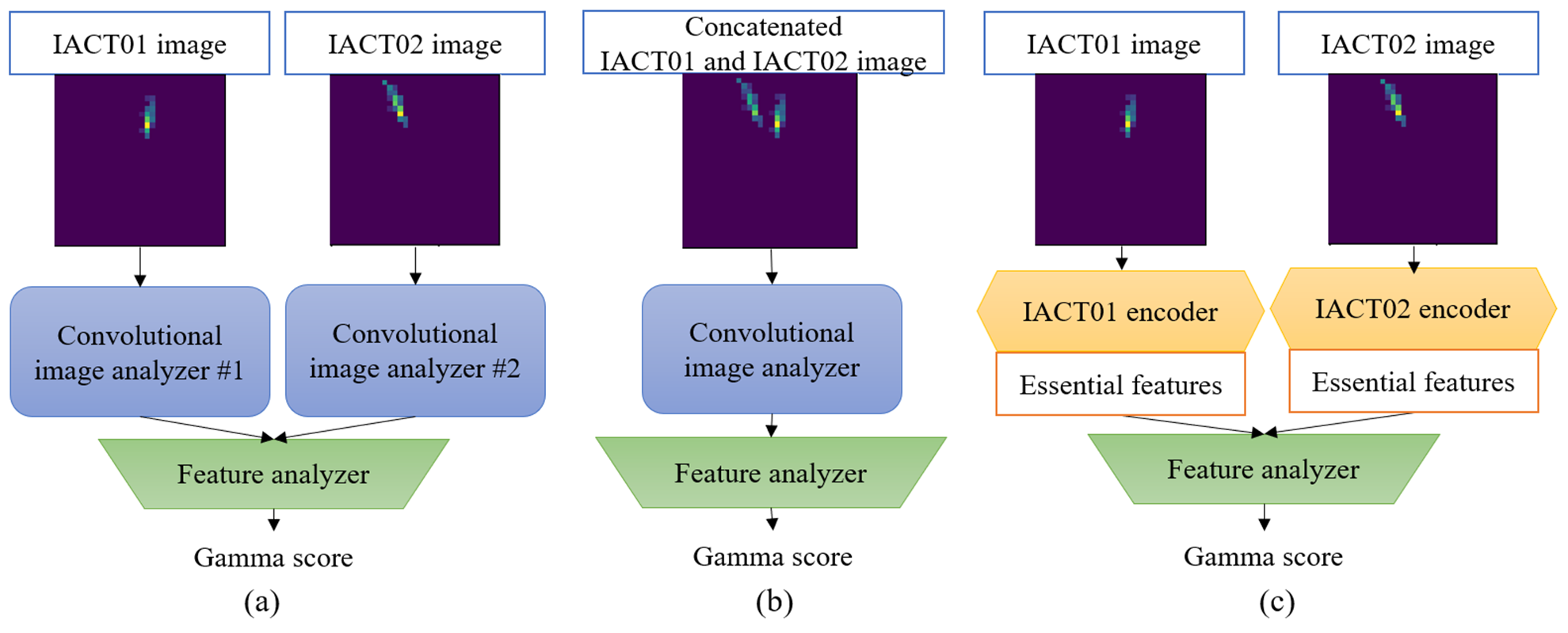}
\caption{\label{fig:fig1}Schematic demonstration for multiple neural network classifier for TAIGA-IACTs images: medium image fusion approach (a),
early image fusion approach (b) and analysis of autoencoder-derived essential features (c).}
\end{figure*}

Several approaches were considered for neural network methods in stereo-mode (Fig.~\ref{fig:fig1}).

The first option involved the analysis of each telescope image separately. This method is called medium fusion, since in this case some of the features extracted by convolutional layers during neural network training are concatenated for further analysis. In the second approach, we proposed to merge IACT images before the input into a neural network model consisting of a convolutional and fully connected analyzer (early fusion). In the third option, we consider the formation of image essential features that are not limited with geometric or statistical considerations, such as the Hillas parameters\cite{5}. For this approach, the autoencoder (AE) pre-trained for each telescope separately is the way to determine the essential features in its latent space\cite{6, 7}. It was shown\cite{8} that in the task of gamma-ray energy reconstruction the analysis of essential features demonstrated the recovery quality no worse than usual convolutional neural networks (CNNs) that processed images directly.

\begin{figure}
\includegraphics{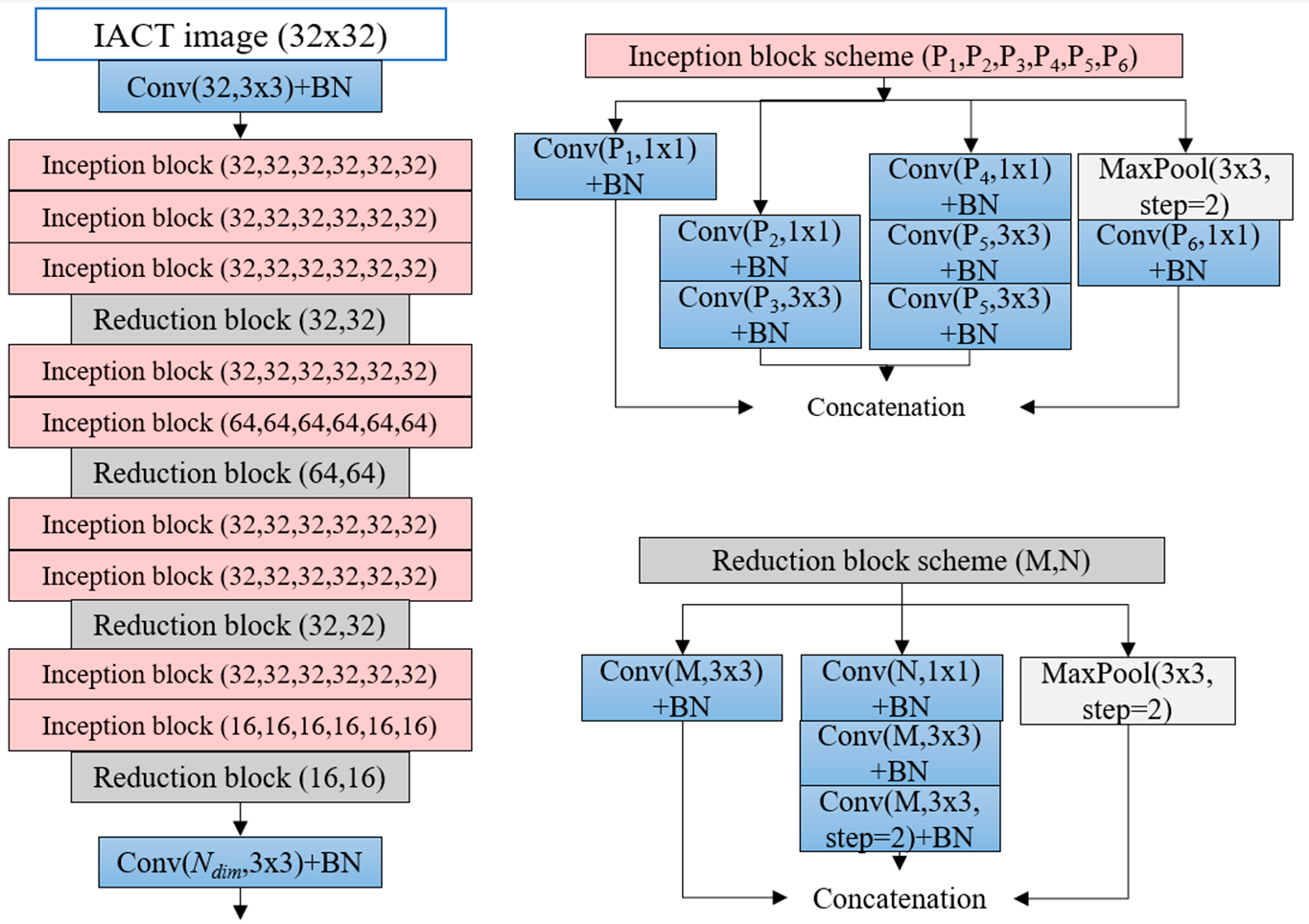}
\caption{\label{fig:fig2}Architecture of convolutional analyzers of medium, early fusion approaches and in AE encoder. \textit{Conv}, \textit{MaxPool} and \textit{BN} are abbreviations for convolutional layer, batch normalization and maximum pooling layer respectively.}
\end{figure}

For the experimental clarity, in all the approaches considered the shape of the convolutional blocks coincided. The architecture of the convolutional block is shown in Fig.~\ref{fig:fig2}, where the dimension of the last convolutional layer was approximately the same as the one of the autoencoder latent space\cite{6}. In the study of autoencoder data compression for TAIGA-IACT images\cite{8} it was found out that the optimal compression quality achieved with the dimension of latent space equals to 12-18. In this work, we have taken the dimension 18. Thus, in the early fusion method, the output layer dimension was 36 because of two telescope images summed up together. In other cases it was 18.

The difference in the three methods affected the structure of the feature analyzer. In the case of medium and early fusion approaches, features extracted with the convolutional block were transmitted to one hidden fully connected layer with 2048 neurons. When analyzing the autoencoder-derived essential features, a more complex structure was used, consisting sequentially of one fully connected layer with 2048 neurons, which was converted into feature maps of sizes (8,8,32). Three convolutional layers with 128, 256 and 512 filters, where the filter size is (2,2) and the step is 2, followed. The last hidden layer was a fully connected layer with 512 neurons. This structure, compared with structures from fully connected layers only, showed the best result in the task of regression and classification from the AE latent space. We considered it impractical to complicate the feature analyzer blocks in the early and medium fusion approaches, since in these cases neural network models extract features, already aiming at the classification task, unlike the situation with the AE.

Thus, the number of weight coefficients for the classifier in the medium fusion approach was approximately 2 million, in the early fusion - 1.1 million, in the AE-derived essential features method - 2.5 million (summed with the encoder). The activation function between the layers was a sigmoid linear function known as SiLU. Two classifier loss functions were considered: binary crossentropy and focal error. The error function of the AEs was mean squared error. Each neural network (NN) output value defined the gamma score, a measure of the neural network confidence that the image under consideration was a gamma-ray event.

\subsection{\label{sec:level2}Training set}

The classifiers were trained on events modeled using the CORSIKA software\cite{9}. The simulation included the generation and calculation of Cherenkov light fluxes during the EAS development in the atmosphere and the modeling of the TAIGA-IACTs technical and geometric characteristics: mirror diffraction, relative PMT sensitivity with Winston cones, background night light noise, etc\cite{10}. Training samples included 30000 gamma ray events and 157000 proton events. To alleviate the training set imbalance, weights for different classes were applied as $(N_{class})^{-1}*(N_{total}/2)$ where $N_{class}$ is the number of events of considered class (gamma or protons) in the training set, $N_{total}$ is an amount of all training samples. In the analysis of essential features, autoencoder was trained on 500000 proton images separately before classifiers. The energy range of every set was 40-400 TeV (with slope -1), zenith angle range was $30-40^\circ$.

Each image has passed through the following preprocessing before input in CNNs. First, cleaning of images was applied to delete non-zero noise pixels with low amplitudes (less than 7 phe). Secondly, every event was scaled logarithmically as $(4.7)^{-1}*log_{10}(x+1)$ where $x$ is pixel amplitude in photoelectrons. Since IACT images have a hexagonal grid, images were transformed to a square grid, using axial method\cite{11}, leading to images with shape of (32,32,1).

\subsection{\label{sec:level2}Validation set}

\begin{figure*}
\includegraphics{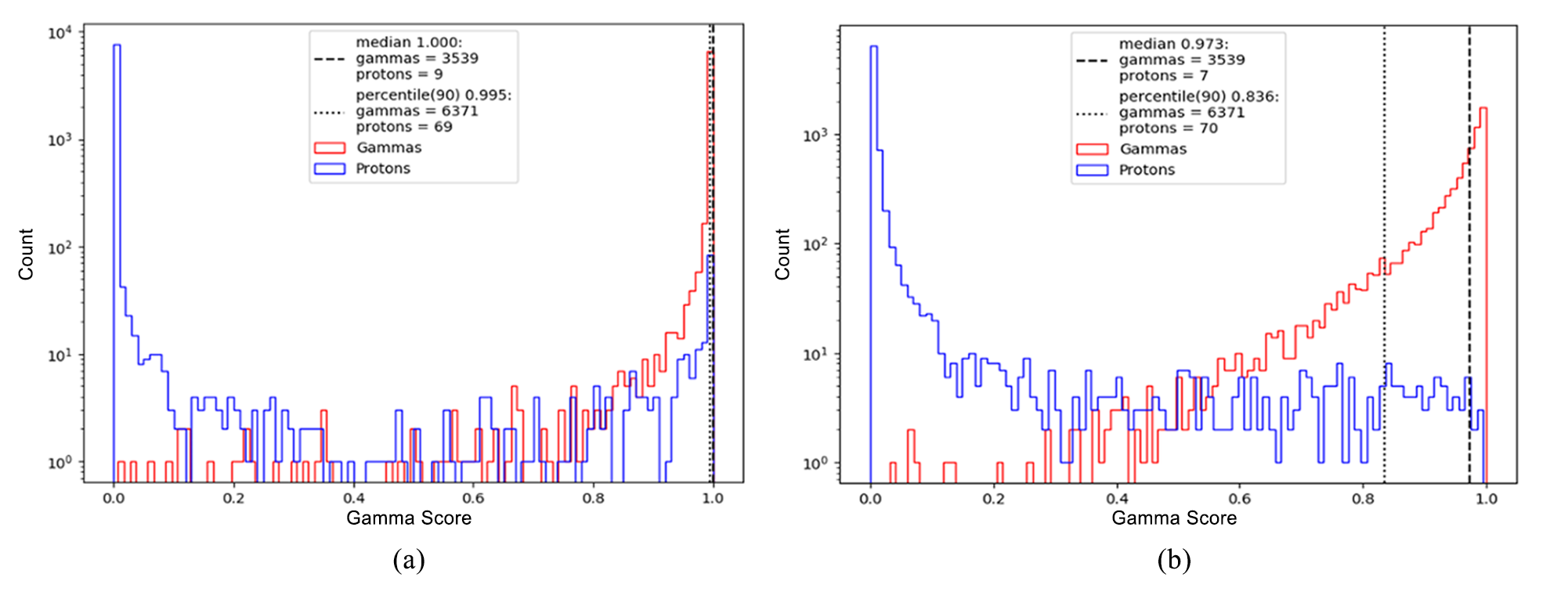}
\caption{\label{fig:fig3}Validation results for the model of the early fusion approach, using binary crossentropy loss function (a) and focal loss function (b).}
\end{figure*}

Using validation samples of 7100 gamma and 7600 proton images, we defined the classification threshold and compared several loss functions for gamma/hadron selection in TAIGA‑IACT.  Figure~\ref{fig:fig3} shows the validation result for classifier based on the early fusion approach. The value when 50\% of gamma events remaining after classification defined our classification threshold. Thus, for crossentropy loss, threshold was equal to 0.99994, while gamma score distribution showed a peak near value 1. Such non-linearity of proton distribution wasn't observed for the training on the focal loss, meaning better class recognition. Also, the behavior of gamma score distribution for other classifier models trained on crossentropy loss hadn't differed a lot from the demonstrated one, so further study was carried out with focal loss function.

\begin{figure*}
\includegraphics{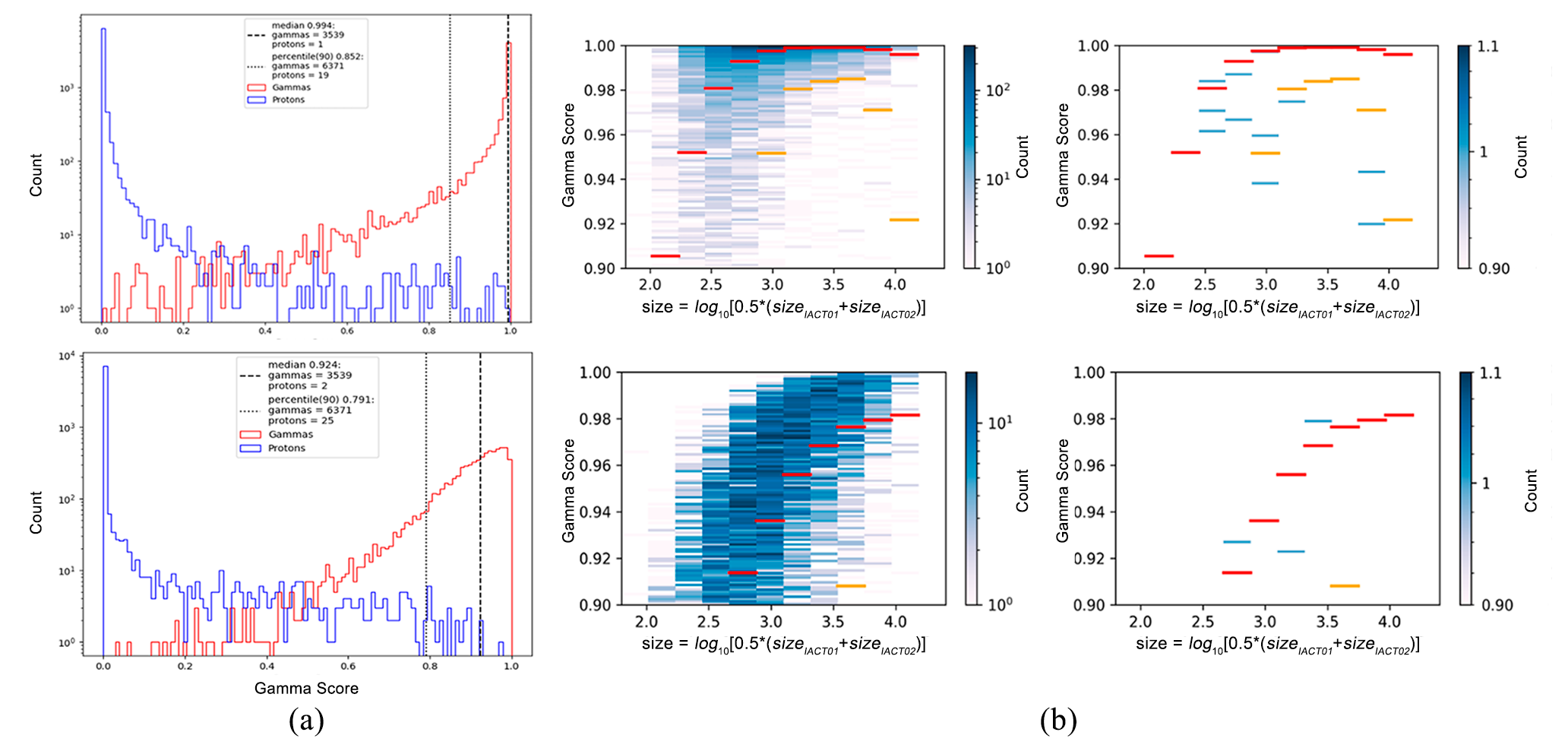}
\caption{\label{fig:fig4} Validation classification results for medium fusion approach (top row) and essential feature analysis (bottom row).
(a) Validation results with the size-independent classification threshold. (b) Dependence of gamma score on image brightness (size). Red - 50\% classification threshold, orange - 90\% threshold of remaining gamma.}
\end{figure*}

Figure~\ref{fig:fig4} demonstrates the comparison of the medium fusion approach with the method of essential feature analysis, including the dependence of gamma score on image brightness known as size\cite{5}. For the early fusion approach distributions are almost equal to the results shown for the medium fusion approach. Distributions clearly show that the essential feature analysis approach has a bad class sensitivity, characterized by the fact that the threshold decreases rapidly at size range less than 500 phe. However, the application of the size-independent threshold (one threshold value for all size ranges) resulted in the almost equal hadron suppression defined ratio of total hadron to hadron passed the selection and classification accuracy and equaled to 7000. AU ROC was 99.8-99.9\% for all neural network methods.

\section{Gamma/hadron suppression in case of strong class imbalance}

For more accurate assessment of neural network methods considered in the previous section, test samples were formed. In experimental observations, significant gamma/hadron radiation imbalance is observed. The ratio of such an imbalance is approximately 1/10000 (for Crab Nebula in TeV energy range). Thus, test set included around 400 gamma EAS images and around 40000 proton images which weren't used in training and validation. The energy range of test samples was 40-400 TeV, zenith angle range was  $30-40^\circ$.

\begin{table}[htbp]
\renewcommand{\arraystretch}{1.25}
\renewcommand{\tabcolsep}{3pt}
\begin{center}\caption{Assessment of the various deep learning classifiers in case of stereoscopic and monoscopic observation with IACTs of TAIGA complex}
\begin{tabular}{|p{1.2in}|p{0.6in}|p{0.6in}|p{1.4in}|p{0.6in}|p{0.6in}|p{0.8in}|} \hline
{Classification threshold} &  {Total gamma} & {Total hadrons}& {Deep learning\newline method} & {Selected gamma} & {Selected protons} & {Hadron suppression} \\ \hline
\multirow{4}{*}{Size-independent} & \multirow{3}{*}{400} & \multirow{3}{*}{39600} &  {Medium fusion} & {188} & {9} & {4400} \\ \cline{4-7}
{}  & {} &  {} &                                            {Early fusion} & {206} & {3} & {13200} \\ \cline{4-7}
{}  & {} &  {} &                                            {Essential feature\newline analysis} & {206} & {3} & {2475} \\ \cline{2-7}
{}  & {320} &  {35000} &  {Mono-mode (CNN)} & {147} & {41} & {854} \\ \hline
\multirow{4}{*}{Size-dependent} & \multirow{3}{*}{400} & \multirow{3}{*}{39600} &  {Medium fusion} & {188} & {26} & {1523} \\ \cline{4-7}
{}  & {} &  {} &                                            {Early fusion} & {203} & {18} & {2200} \\ \cline{4-7}
{}  & {} &  {} &                                            {Essential feature\newline analysis} & {217} & {40} & {990} \\ \cline{2-7}
{}  & {320} &  {35000} &  {Mono-mode (CNN)} & {151} & {119} & {294} \\ \hline
\end{tabular}\label{tab:table1}
\end{center}
\end{table}

Also, for comparison mono-mode was considered as well. In mono-mode, CNN architecture was equal to early fusion approach and was trained only on images from one Cherenkov telescope.
 
Quantitative assessment for test samples are shown in the Table~\ref{tab:table1}. In each case, classification threshold was determined on validation set. As we can see, in both cases (with size-dependent and size-independent classification thresholds) early fusion approach has the highest value of hadron suppression. Also, it can be noticed that the dependency of classification threshold on size parameters doesn't give us advantage in the classification task: a few more gamma events were selected while the false-positive rate (false gamma) increased by at least a factor of 3. So, for classification for very low image sizes (as it was shown in previous section, less than 300 phe) more accurate selection methods are required. In the meantime, assessment demonstrates that deep learning approaches applied in stereo-mode can achieve the desired hadron suppression in comparison to TAIGA-IACT mono-mode.

\section{Conclusion}
In the task of gamma/hadron selection in TAIGA-IACT stereo-mode, multiple deep learning methods were proposed and studied in the
 classification task: medium fusion (separate IACTs images analysis in CNN before concatenation), early fusion (concatenated IACTs 
images are fed to CNN) and autoencoder-derived essential feature analysis. Although autoencoder-derived essential feature analysis is a universal method which can be applied in the processing of different, often incomparable types of data, this method in our task hadn’t reached the desired selection accuracy. Tests results demonstrated that for strongly imbalanced classes the early fusion is the most promising method. In the meantime, with the size-independent classification threshold, classifiers quality significantly improves in comparison to the size-dependent class threshold because a lot of low-intensity EAS images hadn't passed the selection. This factor should be taken into account in the further image analysis concerning the reconstruction of physical parameters after selection with neural network methods. Also, it was demonstrated, that the processing of stereo-mode data doesn't require additional cuts or preselection criteria unlike neural methods for mono-mode IACT data.

\begin{acknowledgments}
The authors would like to express their gratitude 
to the staff of TAIGA collaboration for their support and
recommendations during this work.
\end{acknowledgments}

\section*{FUNDING}
This work was supported by the Russian Science Foundation (grant no. 24-11-00136), using equipment provided under the Lomonosov Moscow State University Development Program and data from the unique TAIGA scientific facility.


\end{document}